\documentclass[prb,twocolumn,showpacs,superscriptaddress]{revtex4}
\usepackage{graphicx,amsmath}

\usepackage{bm}

\usepackage{amsmath}
\usepackage{amsbsy}
\usepackage{amssymb}
\usepackage{amsfonts}

\newcommand{\V}[1]{{\bf #1}} % my-vector
\newcommand{\lav}{\left\langle\!\left\langle}
\newcommand{\rav}{\right\rangle\!\right\rangle}
 
\begin{document}

\title{Coulomb scattering in nitride based self-assembled quantum-dot
       systems}

\author{T. R. Nielsen} 
\email{www.itp.uni-bremen.de}
\affiliation{Institute for Theoretical Physics,  
             University of Bremen,
             28334 Bremen, Germany}

\author{P. Gartner}
\affiliation{Institute for Theoretical Physics,  
             University of Bremen,
             28334 Bremen, Germany}
\affiliation{National Institute for Materials Physics, POB MG-7, 
             Bucharest-Magurele, Romania}

\author{M. Lorke}
\affiliation{Institute for Theoretical Physics,  
             University of Bremen,
             28334 Bremen, Germany}

\author{J. Seebeck}
\affiliation{Institute for Theoretical Physics,  
             University of Bremen,
             28334 Bremen, Germany}

\author{F. Jahnke}
\affiliation{Institute for Theoretical Physics,  
             University of Bremen,
             28334 Bremen, Germany}

\date{\today}


\begin{abstract}
We study the carrier capture and relaxation due to Coulomb scattering
in group-III nitride quantum dots on the basis of population kinetics. For
the states involved in the scattering processes the combined influence
of the quantum-confined Stark effect and many-body renormalizations is
taken into account. The charge separation induced by the built-in
field has important consequences on the capture and relaxation
rates. It is shown that its main effect comes through the
renormalization of the energies of the states involved in the
collisions, and leads to an increase in the scattering efficency. 
\end{abstract}

\maketitle
%%%%%%%%%%%%%%%%%%%%%%%%%%%%%%%%%%%%%%%%%%%%%%%%%%%%%%%%%%%%%%%%%%%%%%%%%%%%%%
\section{Introduction}
In recent years, quantum dots (QDs) have emerged as a powerful tool to
tailor the light-emission properties of semiconductors.\cite{Bimberg:98} 
Applications range from quantum dot lasers and
ultrafast amplifiers to cavity-quantum electrodynamics and
nonclassical light generation.\cite{Michler:03} As a new material
system, group-III nitrides are of intense current interest due to
their extended range of emission frequencies from amber up to
ultraviolet as well as their potential for high-power/high-temperature
electronic devices.\cite{Nakamura:98,Jain:00} While the nonradiative
loss of carriers due to trapping at threading dislocations lowers the
efficiency of group-III nitride quantum-well light emitters, this
effect is reduced by the three-dimensional confinement in 
QDs.\cite{Damilano:99}  Studies of the photoluminescence spectra
\cite{Kalliakos:04} and dynamics \cite{Krestnikov:02,Oliver:03} are
used to demonstrate and analyze efficient recombination processes from
the localized QD states. 

Extensive work has been done to study the electronic states in
group-III nitrides.\cite{Vurgaftman:03} The valence-band structure
has a strongly 
non-parabolic dispersion and a pronounced mass anisotropy. \cite{Chuang:96}
Nitride-based heterostructures with a wurtzite crystal structure are
known to have strong built-in electrostatic fields due to spontaneous
polarization and piezoelectric effects which have been analyzed in
ab-initio electronic structure calculations 
\cite{Bernardini:97,Bernardini:98} and in comparison with
photoluminescence  
experiments.\cite{Cingolani:00} Tight-binding calculations of QD
states have been used to study free-carrier optical
transitions.\cite{Ranjan:03}

While the aforementioned theoretical investigations are devoted to the
single-particle states and transitions, it is known from GaAs-based QD
systems, that the emission properties are strongly influenced by
many-body effects. Self-organized QD systems, grown in the 
Stranski-Krastanov mode,
exhibit a  single-particle energy spectrum with discrete energies for
localized states as well as a quasi-continuum of delocalized wetting
layer (WL) states at higher energies.  The carrier-carrier Coulomb 
interaction provides efficient scattering processes from the 
delocalized into the localized states (carrier capture) as well as 
fast transitions between localized states (carrier relaxation), which
can be assisted by carriers in bound (QD) or extended (WL) states.  
The dependence of the scattering efficiency on the 
excitation conditions has been calculated for GaAs-based QDs on
various levels of 
refinement.\cite{Bockelmann:92,Vurgaftman:94,Brasken:98,Magnusdottir:03,Nielsen:04} 
The scattering rates are of central importance for the
photoluminescence as well as for the laser efficiency and dynamics. The same
interaction processes also lead to a renormalization of the electronic
states (resulting in line shifts for the optical transitions) as well
as to dephasing (line broadening) effects, which directly determine
absorption and gain spectra. \cite{Schneider:04}

With the current attention on 
nitride-based QDs the question raises, to which extent previous
results are modified by the peculiarities of this material
system. Specifically, we study how the strong electrostatic fields and
the corresponding changes of the single-particle wave functions and
energies influence the carrier-carrier scattering
processes. For this purpose, we have to analyze the competing
influence of the internal fields and of the many-body
renormalizations. Since previous 
investigations of carrier-carrier scattering in semiconductor QDs have
been performed for free-carrier energies entering the scattering integrals,
an independent second purpose of this paper is the inclusion of
self-consistently renormalized energies. 

Renormalizations of the single-particle energies are due to direct
electrostatic (Hartree) Coulomb interaction, exchange interaction, 
and screening.   
While these effects generally contribute due to possible charging of
the QDs, in the considered wurtzite structure they are additionally
modified by the presence of the built-in fields.  
In this paper we show that the discussed changes of the
single-particle energies have a much stronger impact on the carrier
scattering processes than modifications of the single-particle wave
functions.

Many-body effects were also considered for the case of few excited
carriers restricted to localized states of GaN-QD.\cite{Rinaldis:04} 
In this regime, the emission properties reflect
the multi-exciton states. In our paper, we are interested in kinetic
processes for the opposite limit of high densities of carriers
populating both QD and WL states (as typical for QD lasers).  

For the evaluation of scattering processes, we use kinetic equations
for the carrier occupation probabilities which include direct and
exchange Coulomb interaction under the influence of carrier screening
as well as population effects (Pauli-blocking) of the involved
electronic states.  Based on the single particle states and Coulomb
interaction matrix elements, scattering processes are evaluated on the
level of second-order Born approximation.\cite{Nielsen:04}

The paper is organized as follows. In Sec.~\ref{sec:theory} we
summarize the main ingredients of our theory which include the kinetic
equations, the single-particle states and their renormalization as
well as the interaction matrix  elements. Details regarding the WL
states and screening effects are given in appendices. In
Sec.~\ref{sec:results} the QD model is described and the numerical
results for energy renormalization and scattering times are presented.

%%%%%%%%%%%%%%%%%%%%%%%%%%%%%%%%%%%%%%%%%%%%%%%%%%%%%%%%%%%%%%%%%%%%%%%%%%%%%%%
\section{Theory for carrier-carrier Coulomb scattering}\label{sec:theory}

To analyze the role of carrier-carrier scattering in nitride-based
QD systems, we use a kinetic equation with Boltzmann scattering
integrals.\cite{Nielsen:04} The dynamics of the carrier population
$f_\nu(t)$ in an arbitrary state $\nu$ is determined by in-scattering
processes, weighted with the nonoccupation of this state, and
by out-scattering processes, weighted with the occupation according to
%%%%%%%%%%%%%%%%
\begin{eqnarray}
  \frac{\partial}{\partial t} \: f_\nu 
  &=&  
  (1-f_{\nu}) S_{\nu}^{\text{in}} - f_{\nu} S_{\nu}^{\text{out}} .
\label{eq:in-out-scatt}
\end{eqnarray}
%%%%%%%%%%%%%%
The in-scattering rate is given by
%%%%%%%%%%%%%%%%
\begin{align}
     S_{\nu}^{\text{in}}
    &= \frac{2 \pi}{\hbar} \sum \limits_{\nu_1,\nu_2,\nu_3} \:
        W_{\nu \nu_2 \nu_3 \nu_1} [ W^*_{\nu \nu_2 \nu_3 \nu_1} 
                                  - W^*_{\nu \nu_2 \nu_1 \nu_3}]
 \nonumber \\
    &
 \times \big\{  f_{\nu_1} (1 - f_{\nu_2}) f_{\nu_3} 
 \delta (\widetilde{\varepsilon}_\nu - \widetilde{\varepsilon}_{\nu_1} 
       + \widetilde{\varepsilon}_{\nu_2} - \widetilde{\varepsilon}_{\nu_3}) 
         \big\} ,
\label{eq:in-rate}
\end{align}
%%%%%%%%%%%%%%
where the first and second term in the square brackets correspond to
direct and exchange Coulomb scattering, respectively.  The matrix
elements of the screened Coulomb interaction, $W_{\nu \nu_2 \nu_3
\nu_1}$, are calculated in Section~\ref{sec:CoulMartix}.  The
delta-function describes energy conservation in the Markov limit.
Single-particle energies $\widetilde{\varepsilon}_\nu$ of state $\nu$
are renormalized by the Coulomb interaction as discussed in
Sections~\ref{sec:HF} and \ref{sec:CH}.  A similar expression for the
out-scattering rate $S_{\nu}^{\text{out}}$ is obtained by replacing $f
\rightarrow 1-f$.

Scattering processes described by Eqs.~\eqref{eq:in-out-scatt} and
\eqref{eq:in-rate} conserve the total energy and separately the
electron and hole numbers. As a result, the combined action of the
discussed scattering processes will evolve the distribution functions
of electrons and holes towards Fermi-Dirac functions with common
temperature where the QD and WL electrons (holes) will have the same
chemical potential $\mu_e$ ($\mu_h$).  During such a time evolution
towards quasi-equilibrium the relative importance of various scattering
processes is expected to change via their dependence on the
(non-equilibrium) carrier distribution functions for WL and QD states.
A direct measure for the efficiency of various scattering processes can
be given in the relaxation-time approximation \cite{Nielsen:04} where one
calculates the characteristic time on which a small perturbation of
the system from thermal equilibrium is obliterated.  We use this
method to investigate the influence of the built-in electrostatic
field in nitride-based QDs on the carrier-density dependent scattering
efficiency.

%%%%%%%%%%%%%%%%%%%%%%%%%%%%%%%%%%%%%%%%%%%%%%%%%%%%%%%%%%%%%%%%%%%%%%%%%%%%%%%
\subsection{Quantum-dot model system} \label{sec:QDmodel}

Recent progress in tight-binding and $\V{k}\cdot\V{p}$ models has been made in
calculating QD electronic single-particle states including 
the confinement geometry, strain and built-in electrostatic field effects.
The special features of the wurtzite QD structures have been addressed
in Refs.~[\onlinecite{Ranjan:03,Andreev:00,Bagga:03,Fonoberov:03}], while
zinc-blende QD structures have been studied in 
Refs.~[\onlinecite{Bagga:03,Fonoberov:03,Sheng:05,Williamson:00,Bester:05,Schulz:05xxx}].
Our goals differ from these investigations
in the respect that, for given single-particle states and energies, we
need to determine Coulomb interaction matrix elements in order to
calculate many-body energy renormalizations and scattering
processes. For this purpose we choose a simple representation of wave
functions for lens-shaped quantum dots \cite{Wojs:96} which allows to
separate the in-plane motion (with weak confinement for the states
localized at the QD position and without confinement for the states
delocalized over the WL plane) from the motion in growth direction
with strong confinement. This leads to the ansatz
%%%%%%%%%%%%%%%%
\begin{equation}
\Phi_{\nu}({\bf r})=\varphi_l^{b}({\bm\varrho}) \; 
                    \xi_\sigma^{b}(z) \; u_{b}({\bf r})
\label{eq:wf1}
\end{equation}
%%%%%%%%%%%%%%%%
where the WL extends in the plane described by $\bm\varrho =(x,y)$.
$\varphi$ and $\xi$ are the envelope functions in this plane and in
the perpendicular growth direction, respectively, and $u$ are Bloch
functions.  
$\nu$ represents a set of quantum numbers with $l$ for the in-plane
component (including the spin), 
$\sigma$ for the $z$-direction, and $b$ is the band index.
In the following we consider an ensemble of
randomly distributed identical QDs with non-overlapping localized
states. The total number of QDs, $N$, leads in the large area
limit to a constant QD density $n_{\text{QD}}= \lim_{A \rightarrow
\infty} N/A$.

Regarding the dependence of the results on the choice of
wave functions, we find that not so much the particular
form but rather the correct symmetry is of relevance.
Even tough a more accurate treatment is expected to mix the in-plane and 
$z$-coordinates, the ansatz \eqref{eq:wf1} is preserving the symmetry.

%%%%%%%%%%%%%%%%%%%%%%%%%%%%%%%%%%%%%%%%%%%%%%%%%%%%%%%%%%%%%%%%%%%%%%%%%%%%%%%
\subsection{Coulomb matrix elements} \label{sec:CoulMartix}

The interaction matrix elements of the bare Coulomb potential 
$v({\bf r}-{\bf r'})=e^2/(4\pi\varepsilon_0 \epsilon |{\bf r}-{\bf r'}|)$ 
with the background dielectric function $\epsilon$ are given by
%%%%%%%%%%%%%%%%
\begin{align}
    V_{\nu \nu_2 \nu_3 \nu_1} \nonumber & =  
    \int d^3 r \; d^3 r' \;
\nonumber \\ &
    \Phi^*_{\nu}({\bf r}) \Phi^*_{\nu_2}({\bf r'})
    v({\bf r}-{\bf r'})
    \Phi_{\nu_3}({\bf r'}) \Phi_{\nu_1}({\bf r}) .
\label{eq:Coulomb}
\end{align}
%%%%%%%%%%%%%%%%
This expression is further specified with the help of Eq.~\eqref{eq:wf1},
the Fourier transform of the Coulomb potential, and by introducing the 
in-plane Coulomb matrix elements with the two-dimensional
momentum ${\bf q}$,
%%%%%%%%%%%%%%%%
\begin{align}
    V_{\sigma \sigma_2 \sigma_3 \sigma_1}^{b,b'}({\bf q}) 
    & = \frac{e^2}{2\varepsilon_0\epsilon \, q}
    \int dz \; dz' \;
\nonumber \\ &
    \xi_\sigma^{b}(z)^* \xi_{\sigma_2}^{b'}(z')^*
    e^{-q |z-z'|}
    \xi_{\sigma_3}^{b'}(z') \xi_{\sigma_1}^{b}(z) .
\label{eq:Coulomb1}
\end{align}
%%%%%%%%%%%%%%
Limiting the calculations to the first bound state of the strong
confinement problem (in $z$-direction), all the $\sigma$ indices above take
only one value and will be dropped in what follows.
Also, the band indices associated with $\nu_1$ and $\nu_3$ are the same
as those of $\nu$ and $\nu_2$, respectively, so that only two of them 
have to be specified.  
Therefore Eq.~(\ref{eq:Coulomb}) reads 
%%%%%%%%%%%%%%%%
\begin{eqnarray}
    V_{l,l_2, l_3, l_1}^{b,b_2} &=& \frac{1}{A} \: \sum \limits_{\bf q} \, 
       V^{b,b_2}({\bf q}) \, 
\nonumber \\
    && \times \int d^2 \varrho \: \varphi_l^{b}({\bm \varrho})^*
       \varphi_{l_1}^{b}({\bm \varrho})  \: e^{-i {\bf q} \cdot
       {\bm \varrho}} \;
\nonumber \\
    && \times   \int d^2 \varrho' \: \varphi_{l_2}^{b_2}({\bm \varrho}')^*
       \varphi_{l_3}^{b_2}({\bm \varrho}') \: e^{i {\bf q} \cdot
       {\bm \varrho}'}.
\label{eq:Coulomb2} 
\end{eqnarray}
%%%%%%%%%%%%%%
The obtained separation of integrals over in-plane and $z$-components
greatly simplifies the computational effort.  For practical
calculations we use solutions of a two-dimensional harmonic potential
for the in-plane wave functions of the localized states and 
orthogonalized plane wave
(OPW)
solutions, discussed in Appendix~\ref{app:a}, for the in-plane
components of the delocalized states. Then the in-plane integrals in
Eq.~\eqref{eq:Coulomb2} can be determined to a large extent
analytically for all possible combinations of QD and WL states. The
calculation of the wave functions in growth direction, which is used
to evaluate the in-plane Coulomb matrix elements,
Eq.~\eqref{eq:Coulomb1}, is outlined in the next section. Screened
Coulomb matrix elements are obtained from Eqs.~\eqref{eq:Coulomb1} and
\eqref{eq:Coulomb2} according to the procedure described in detail in
Ref.~[\onlinecite{Nielsen:04}].

%%%%%%%%%%%%%%%%%%%%%%%%%%%%%%%%%%%%%%%%%%%%%%%%%%%%%%%%%%%%%%%%%%%%%%%%%%%%%%%
\subsection{Quantum-confined Stark effect} \label{sec:SPE}

To account for the built-in electrostatic fields of the wurtzite
structure in the growth-direction, which gives rise to the 
quantum-confined Stark effect, we solve the one-dimensional
Schr\"odinger equation \cite{Chow:99b,Chow:99} 
%%%%%%%%%%%%%%
\begin{eqnarray}\label{eq:1dsg}
  \left[ 
	-\frac{\hbar^2}{2m_{b}} \frac{\partial^2}{\partial z^2} 
        + U^{b}(z)
 \right] \xi^{b}(z)
  = 
 E_{b}(z)\xi^{b}(z) ,
 \end{eqnarray}
%%%%%%%%%%%%%%
where the potential
%%%%%%%%%%%%%%
\begin{eqnarray}\label{eq:1dpot}
U^{b}(z)= U_{0}^{b}(z) + U_{\text{p}}^{b}(z) + U_{\text{scr}}^{b}(z)  ,
\end{eqnarray}
%%%%%%%%%%%%%
consists of the bare confinement potential in $z$-direction,
$U_{0}^{b}(z)$, as well as the intrinsic electrical and screening fields,
$U_{\text{p}}^{b}(z)$ and $U_{\text{scr}}^{b}(z)$, respectively. The
latter is the electrostatic field due to the separation of electron
and hole wave functions by the built-in field.
Following Ref.~[\onlinecite{Chow:99}] we calculate the corresponding screening 
potential from a solution of the Poisson equation for a set of uniformly
charged sheets  according to
%%%%%%%%%%%%%
\begin{eqnarray}\label{eq:1dscr}
 U_{\text{scr}}^{e,h}(z) =\frac{\mp e^{2} N_{\text{sys}}}{2 \varepsilon_0 \epsilon}
 \int \mathrm{d}z' 
 \left[ 
         |\xi^{e}(z')|^2 - |\xi^{h}(z')|^2
 \right]    
 |z-z'| .
\end{eqnarray}
%%%%%%%%%%%%%
Equations \eqref{eq:1dsg}-\eqref{eq:1dscr} have to be evaluated
selfconsistently for a given total (QD plus WL) carrier density 
$N_{\text{sys}}$ in the system.

%%%%%%%%%%%%%%%%%%%%%%%%%%%%%%%%%%%%%%%%%%%%%%%%%%%%%%%%%%%%%%%%%%%%%%%%%%%%%%%
\subsection{Hartree-Fock energy renormalization} \label{sec:HF}

The Hartree-Fock (HF) contribution to the energy renormalization of an
arbitrary state $\nu$ (QD or WL) is given by
%%%%%%%%%%%%%%%%
\begin{eqnarray}
  \widetilde{\varepsilon}_{\nu} & = &  \label{eq:energy_hf}
  \varepsilon_{\nu}  + \Delta^{\text{HF}}_{\nu} ,
\end{eqnarray}
%%%%%%%%%%%%%%%%
where $\varepsilon_{\nu}$ is the free-carrier energy and the HF shift
follows from
%%%%%%%%%%%%%%%%
\begin{eqnarray}
 \Delta^{\text{HF}}_{\nu} & = &   \label{eq:sigma_hf}
 \Delta^{\text{H}}_{\nu}  + \Delta^{\text{F}}_{\nu}  \\
 & = & 
 \sum_{\nu'} 
 \left[ 
         V_{\nu \nu' \nu' \nu} 
       - V_{\nu \nu' \nu \nu'} 
 \right]
 f_{\nu'} . \notag
\end{eqnarray}
%%%%%%%%%%%%%%%%%
The first part corresponds to the Hartree (direct) term and the second
part is the Fock (exchange) contribution. The equation is written in a
general basis. For a system with local charge neutrality, like bulk
semiconductors or quantum wells, the Hartree term vanishes while the
Fock term leads to an energy reduction as used, e.g., in the
semiconductor Bloch equations.\cite{Haug:94}

For the QD and OPW-WL states discussed in this paper, the absence of
local charge neutrality leads to Hartree terms which
are evaluated in Appendix~\ref{app:b}.  Regarding the quantum numbers,
introduced in Section~\ref{sec:QDmodel}, we further specify the
following notation: For the in-plane envelopes we use the two-dimensional
momentum $\V{k}$ for the delocalized WL states and $\alpha=(m,\V{R})$
for the localized QD states, where $\V{R}$ is the QD position and the
discrete quantum numbers for a particular QD are collected in $m$. The
spin index is tacitly included in either $m$ or $\V{k}$. 

Following Appendix~\ref{app:b}, we find that the Hartree shifts of the
WL states vanish due to 
compensating contributions from QD and WL carriers,
%%%%%%%%%%%%%%%%%
\begin{eqnarray}
 \Delta^{\text{H}}_{b, \V{k}} & = & 0 .
\label{eq:H_wl}
\end{eqnarray}
%%%%%%%%%%%%%%%%%
For a random distribution of QDs this is related to the spatial
homogeneity restored on a global length scale and to the global
charge neutrality of the system. For the same reason, 
the Hartree shift of a localized QD state 
is only provided by 
states from the same QD
while Hartree shifts due to carriers in other QD and WL
states compensate each other,
%%%%%%%%%%%%%%%%
\begin{eqnarray}
  \Delta^{\text{H}}_{b, m} & = &
  \sum_{b',m'} \;
  V_{m m' m' m}^{b,b'} \; 
  f^{b'}_{m'}	\, .
\label{eq:H_dot}
\end{eqnarray}
%%%%%%%%%%%%%%%%%

From electrostatics one expects the same result, provided that the WL
is modeled as a constant area charge of opposite sign to the QD
total charge. Then the constant part of the Fourier expansion of the
Coulomb matrix elements in Eq.~\eqref{eq:Coulomb2}, i.e. the $\V{q}=0$
contribution, for the QDs balances the constant area charge from
the WL.

On the other hand, since $\Delta^{\text{H}}_{b, m}$ probes the
local charge density at the site of the QD due to the contributions of
QD and WL carriers, one expects an influence of the WL on the QD
Hartree energy shift. Locally on the QD length scale, the WL states
are not homogeneous as a result of the QD presence. This causes a
departure from the picture where the WL states contribute only in an
averaged manner (via the $\V{q}=0$ term).  Intuitively, one would
expect that an increasing amount of carriers in the WL will start to
screen the Coulomb interaction between the QD carriers.  Following
this picture we therefore replace the bare Coulomb potential with the
screened one in Eq.~\eqref{eq:H_dot}, i.e.,
$\Delta^{\text{H}}_{b,m} \rightarrow \Delta^{\text{SH}}_{b,m}$. 
In Appendices ~\ref{app:b} and
~\ref{app:c} more support is given to this argumentation.

In the exchange terms, the summation over the QD positions can be
performed directly, since the associated QD phase factors disappear
for the Coulomb matrix elements $V_{m m' m m'}$ and $V_{\alpha \V{k}
\alpha \V{k}}$ as seen from Eq.~\eqref{eq:dd} and Eq.~\eqref{eq:dk},
respectively.  The resulting exchange energy shifts contain the QD and
WL contributions,
%%%%%%%%%%%%%%%%%
\begin{eqnarray}
 \Delta^{\text{F}}_{b, m} = & - & \sum_{m'} \;  \label{eq:F_dot}
  V_{m m' m m'}^{b,b} \,    
 f^{b}_{m'}    \notag \\ \notag \\
 & - &  \sum_{\V{k}'} \;
 V_{m \V{k}' m \V{k}'}^{b,b} \, 
 f^{b}_{\V{k}'}	, \\
 \Delta^{\text{F}}_{b, \V{k}} = & - & N \cdot \sum_{m'} \; \label{eq:F_wl}
 V_{\V{k} m' \V{k} m'}^{b, b} \,
 f^{b}_{m'} \notag \\ \notag \\
 & - &  \sum_{\V{k}'} \;
 V_{\V{k} \V{k}' \V{k} \V{k}'}^{b,b} \, 
 f^{b}_{\V{k}'}  .
\end{eqnarray}
%%%%%%%%%%%%%%%%
Since there is an area associated with the Coulomb matrix element
$V_{\V{k}m'\V{k}m'}$, the QD density $n_{\text{QD}}=N/A$ enters in
Eq.~\eqref{eq:F_wl}.

%%%%%%%%%%%%%%%%%%%%%%%%%%%%%%%%%%%%%%%%%%%%%%%%%%%%%%%%%%%%%%%%%%%%%%%%%%%%%%%
\subsection{Screened exchange and Coulomb hole} \label{sec:CH}

The HF Coulomb interaction provides only the first approximation for
energy renormalizations; correlation contributions can lead to
important corrections. A frequently used extension of the HF energy
shifts for high-density plasma excitation is the screened-exchange and
Coulomb-hole approximation.\cite{Haug:94} On this level, the combined
contributions of Coulomb exchange interaction and Coulomb correlations
beyond HF to the energy renormalizations are approximated with the
screened exchange term (where the bare Coulomb potential in the Fock
term is replaced by a screened one) plus an energy shift denoted as
Coulomb-hole contribution. The  Coulomb hole self-energy
reads \cite{Haug:94}
%%%%%%%%%%%%%%%%
\begin{align}
 \Sigma^{\text{CH}}(\V{r}_1, \V{r}_2,t_1,t_2) & = 
  \frac{1}{2} \delta(\V{r}_1- \V{r}_2)\delta(t_1-t_2)  \notag \\
 & \times
 \left[
 W(\V{r}_1, \V{r}_2,t_1) - V(\V{r}_1- \V{r}_2)
 \right] ,
\end{align}
%%%%%%%%%%%%%%%%
with the statically screened Coulomb potential $W$. In a general
eigenfunction basis, this leads to the Coulomb-hole energy shift
%%%%%%%%%%%%%%%%
\begin{align}
  \Delta^{\text{CH}}_{\nu}(t)  & = 
 \frac{1}{2} \sum_{\nu'}
\left[
       W_{\nu \nu' \nu \nu'}(t) - V_{\nu \nu' \nu \nu'}
\right]  .
\end{align}
%%%%%%%%%%%%%%%%

In Eqs.~\eqref{eq:F_dot} and \eqref{eq:F_wl} we therefore replace the
bare Coulomb potential with the screened one and substitute the
Fock energy shift with the screened exchange plus the Coulomb hole,
$\Delta^{\text{F}}_{\nu} \rightarrow \Delta^{\text{SX}}_{\nu}$ + $
\Delta^{\text{CH}}_{\nu}$.  In the limit of low carrier densities, the
screened Coulomb potential reduces to the bare one, the Coulomb hole
vanishes, and we recover the HF result.

%%%%%%%%%%%%%%%%%%%%%%%%%%%%%%%%%%%%%%%%%%%%%%%%%%%%%%%%%%%%%%%%%%%%%%%%%%%%%%%
\section{Results} \label{sec:results}

We start this Section by describing the model we use and the
material parameters employed for calculating the one-particle
states. These correspond to the unexcited system. The next step is 
to renormalize these states in a self-consistent way to include
the influence of the carrier population and the presence of the built-in 
field, as described in Section ~\ref{sec:theory}. These single-particle 
properties enter the scattering integrals, Eq.~\eqref{eq:in-rate}.

For the discussion of the numerical results one should bear in mind the two 
main 
consequences of the built-in field. 
On the one hand, the self-consistent energies entering the Fermi
functions and the energy-conservation are sensitive to the 
electron-hole separation induced by the built-in electrostatic 
field. On the other hand, the Coulomb matrix elements are
changed due to modifications in the wave-function shapes 
and overlapping. 

%%%%%%%%%%%%%%%%%%%%%%%%%%%%%%%%%%%%%%%%%%%%%%%%%%%%%%%%%%%%%%%%%%%%%%%%%%%%%%%
\subsection{The model and its  parameters}
%%%%%%%%%%%%%%
\begin{table}[tb]%[H]
\begin{ruledtabular}  
\caption{Material parameters used in the calculations.\label{tab:parameters} }
\begin{tabular}{lcccccc}
Parameter		&& GaN				&& InN				&& In$_{0.2}$Ga$_{0.8}$N  		\\
\hline									  				\\
$E_g$[300 K]  (eV)	&& 3.438\footnotemark[1]	&& 0.756\footnotemark[1]	&&  2.677	 \\
$\Delta E_e$ (eV)	&&  				&&    				&&  0.457	\\
$\Delta E_h$ (eV)	&&  				&&    				&&  0.304	\\
$\epsilon$		&& 8.9\footnotemark[2]		&& 15.3\footnotemark[2]		&&  10.2      	\\
$m_e$ ($m_0$)		&& 0.2\footnotemark[1]		&& 0.07\footnotemark[1]		&&  0.174    \\
$A_1$ 			&&-7.21\footnotemark[1]		&&-8.21\footnotemark[1]		&&  -7.41	\\
$A_2$ 			&&-0.44\footnotemark[1]		&&-0.68\footnotemark[1]		&&  -0.488	\\
$A_3$ 			&& 6.68\footnotemark[1]		&& 7.57\footnotemark[1]		&&  6.858	\\
$A_4$ 			&&-3.46\footnotemark[1]		&&-5.23\footnotemark[1]		&&  -3.814	\\
$A_5$ 			&&-3.40\footnotemark[1]		&&-5.11\footnotemark[1]		&&  -3.742	\\
$\Delta_{\text{so}}$ (eV)&& 0.017\footnotemark[1]	&& 0.005\footnotemark[1]	&&   0.0146	\\
Quantum well (nm)               && 				&&				&&  3.0         \\
$F_{\text{QD-WL}}$ (MV/cm)	&&				&&				&&  1.5 	\\ 
$F_{\text{barrier}}$  (MV/cm)	&&				&&				&&  0.75  
\end{tabular}
\footnotetext[1]{ From Ref.~[\onlinecite{Vurgaftman:03}]. }
\footnotetext[2]{ From Ref.~[\onlinecite{Levinshtein:01}]. }
\end{ruledtabular}
\end{table}
%%%%%%%%%%%%%% 
  
In the following examples we consider an InGaN/GaN QD-WL system using 
typical InGaN parameters \cite{Vurgaftman:03} listed in
Table~\ref{tab:parameters}. For the alloy, we have interpolated
linearly the  
dielectric constant $\epsilon$, the isotropic electron mass $m_e$, the 
hole mass parameters $A_{i}$ and the spin-orbit splitting $\Delta_{\text{so}}$.

A specific feature of the wurtzite
structure nitrides is the strong mass anisotropy of the holes. The
mass of the heavy hole (HH) in the $z$-direction, is given 
by $m^{\ast}_{z}=m_0/|A_1+A_3|$  with the free electron mass
$m_0$ and the mass parameters $A_1$ and $A_3$.\cite{Chuang:96} For the
in-plane motion, a strong hybridization between the HH and the
light-hole (LH) subband leads to nonparabolic bands. This is due to the 
fact that the small HH-LH splitting, induced by the spin-orbit 
interaction, is enhanced by neither the strain nor the
$z$-confinement. To include the hybridization effect we use a 
nonparabolic HH dispersion,\cite{Chuang:96}
%%%%%%%%%%%%%%%%
\begin{align}
\label{eq:HH_disp}
 \varepsilon_{\V{k}}^{h} & =  
 -\alpha (A_2 + A_4) \V{k}^2 
 -\sqrt{ \Delta_{2}^2 + \alpha^2 A^{2}_{5}\V{k}^4 } , 
\end{align}
%%%%%%%%%%%%%%%%
where $\alpha= \hbar^2 / 2m_0$ and  $\Delta_{2}=\Delta_{\text{so}}/3$.  
For the motion of the electrons, isotropic effective mass and
parabolic dispersion are 
considered.
Conduction ($\Delta E_e$) and valence ($\Delta E_h$) band offsets for InGaN 
grown on GaN have been estimated by splitting the gap energy difference 
between bulk GaN and the alloy with a band offset ratio 60:40 for electrons 
and holes, according to Refs.~[\onlinecite{Wu:03,Zhang:04}].

Compared to usual zinc-blende structure, the wurtzite structure is 
characterized 
by large built-in electric fields. The strength of the fields is related to the
spontaneous polarization discontinuity at the heterojunction interfaces and 
the piezoelectric polarization.\cite{Bernardini:97,Bernardini:98} Internal 
fields in InGaN/GaN heterojunctions of a few MV/cm have been 
reported,\cite{Lai:02,Zhang:04} typically in a sawtooth profile, where the 
field in the 
QD-WL system has a different magnitude and opposite direction to the field of 
the barrier (reflecting a set of capacitors with non-equal surface charges). 
For the field inside the QD-WL region and in the barrier, we consider
$F_{\text{QD-WL}} =1.5$ MV/cm and $F_{\text{barrier}}=0.75$  MV/cm,
respectively. 

%%%%%%%%%%%%%%
 \begin{table}[tb]
 \begin{ruledtabular}
 \caption{QD parameters used in the calculations.\label{tab:parameters-qd} }
 \begin{tabular}{lcccccc}
 Parameter			&& Electrons		&& Holes	&&	\\
 \hline								&&	\\
 Shells				&& s, p			&&  s, p, d	&&	\\
 Level spacing (meV)		&& 90.0			&& 30.0		&&	\\
 $\varepsilon_{\text{s}}$ (meV)	&& -160.0		&& -80.0	&& 	\\
 $\varepsilon_{\text{p}}$ (meV)	&& -70.0		&& -50.0	&&	\\		
 $\varepsilon_{\text{d}}$ (meV)   	&&			&& -20.0	&&	\\	
 QD density (cm$^{-2}$)		&&			&&		&&	$10^{10}$
 \end{tabular}
 \end{ruledtabular}
 \end{table}
%%%%%%%%%%%%%%

As the estimated effective hole masses \cite{Chuang:96} are 
larger than the electron mass, we expect the QD's to confine
more states for holes than for electrons. On the same basis, the level
spacing which scales inversely with the mass, is larger for electrons
than for holes. 
The in-plane QD confinement is modeled with a 2D parabolic potential,
capable of binding two energy shells  (s and p) for electrons and
three  (s,p,d) for holes. The degeneracies of these shells (apart
from spin) are $1,2$ and $3$ for s,p and d, respectively.
For electrons we assume a level spacing of 90 meV with the p-shell 70 meV 
below the WL continuum edge, while for holes we 
assume a level spacing of 30 meV with the d-shell 20 meV 
below the WL continuum edge. Thus the $e_s - h_s$ QD transition is
close to the range given in
Ref.~[\onlinecite{Oliver:03,Lin:02}]. The 
QD parameters are summarized in Table~\ref{tab:parameters-qd}.
Finally, we assume  $z$-direction confinement wave functions
which are band dependent but equal 
for QD and WL states.\cite{Nielsen:04}

The harmonic oscillator (HO) inverse localization length $\beta$ is
deduced from the level spacing via  $\hbar \omega_{\text{HO}}= \hbar^2
\beta^2 / m^{\ast}$. For electrons the effective mass of Table
\ref{tab:parameters} is taken. For holes we use the mass resulting
from Eq.~\eqref{eq:HH_disp} in the small $k$ limit, $m^{\ast}_{h}
= m_0/|A_2+A_4|$. This is justified by the typical QD diameters of 
100\AA $\;$ - 200\AA  $\;$, which correspond to the region
around $k = 0$ where the first term of Eq.~\eqref{eq:HH_disp} 
is dominant.

%%%%%%%%%%%%%%%%%%%%%%%%%%%%%%%%%%%%%%%%%%%%%%%%%%%%%%%%%%%%%%%%%%%%%%%%%%%%%%%
\subsection{Schr\"odinger and Poisson equations}\label{sec:S_P}

\begin{figure}[tb]
  \begin{center}
  \includegraphics*[width=0.45\textwidth]{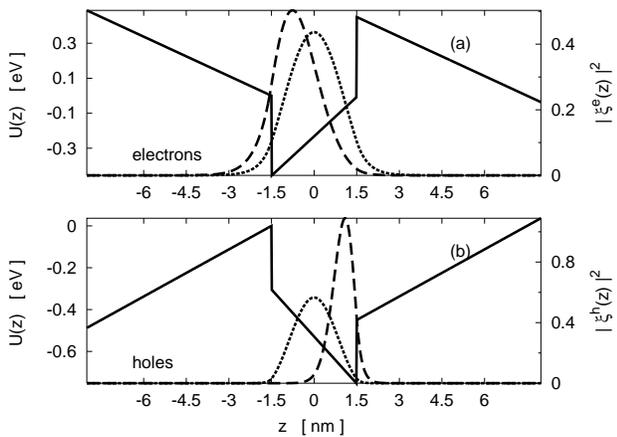}
    \caption{Total confinement potential $U(z)$ (solid line) and modulus 
	 	amplitude of the $z$-component 
                wave function  $|\xi(z)|^{2}$ (dashed 
                line) along the growth direction $z$ for electrons (a) and 
		holes (b) in the presence of the  built-in field at a total
		carrier density $N_{\text{sys}}=10^{10}$ cm$^{-2}$. 
		The dotted lines are the wave function amplitudes for zero 
		field.
             }
    \label{fig:pot-wf}
  \end{center}
\end{figure}

The charge separation of carriers along the growth direction $z$
under the influence of the built-in
electrostatic fields, obtained from a self-consistent solution of
Eqs.~\eqref{eq:1dsg}-\eqref{eq:1dscr} is shown in Fig.~\ref{fig:pot-wf} (a)
and (b). A clear overlap reduction of the wave functions for electrons
and holes is observed. This leads to a
decreasing form factor (the double integral in Eq.~\eqref{eq:Coulomb1}
which modifies the 2D Fourier transform of the Coulomb interaction)
for the electron-hole interaction while the stronger carrier 
localization increases the form factors for electron-electron and
hole-hole interaction. Thus, the presence of the built-in field 
leads to an effective reduction of the electron-hole interaction, 
while the electron-electron and the hole-hole interaction is 
enhanced. The screening field turns out to be a small correction 
to the strong built-in field for the range of densities considered.

%%%%%%%%%%%%%%

%%%%%%%%%%%%%%%%%%%%%%%%%%%%%%%%%%%%%%%%%%%%%%%%%%%%%%%%%%%%%%%%%%%%%%%%%%%%%%%
\subsection{Renormalized energies}\label{sec:ren}

%%%%%%%%%%%%%%
\begin{figure}[tb]
  \begin{center}
  \includegraphics*[width=0.45\textwidth]{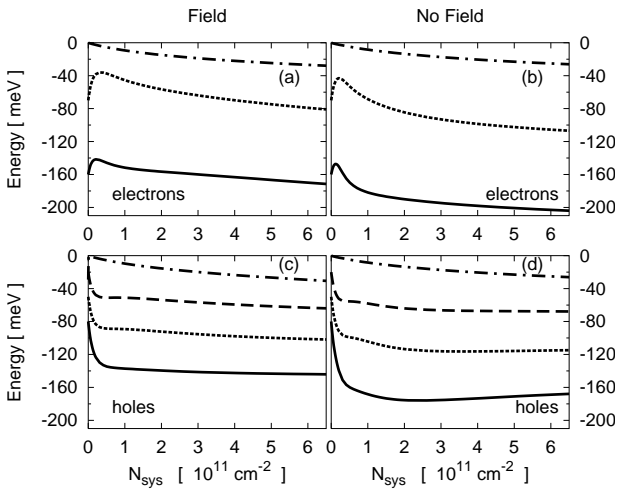}
    \caption{Renormalized  energies as a function of the total carrier
             density in system $N_{\text{sys}}$ for s-shell (solid lines), 
		p-shell (dotted lines), 
		d$_\pm$-shell (dashed lines),
		and WL $\V{k}=0$ (dashed-dotted lines). 
		 Calculation with (a),(c) and
             	without (b),(d) electrostatic field are shown for electrons 
             	(a),(b) and holes (c),(d).  The temperature is 300 K.
             }
    \label{fig:QD-energy}
  \end{center}
\end{figure}

For a total carrier density $N_{\text{sys}}$, the corresponding
carrier distributions $f_{\nu}$ in thermal equilibrium are used to
determine the renormalized energies from the self-consistent
solution of the equation 
$\widetilde{\varepsilon}_{\nu}= \varepsilon_{\nu} +
  \Delta_{\nu}^{\text{SH}} +
  \Delta_{\nu}^{\text{SX}} +
 \Delta_{\nu}^{\text{CH}}$, 
for the QD and WL states. 
For $\varepsilon_\nu $ we use the energies listed in
Table~\ref{tab:parameters-qd} for the QD bound states and the
dispersion law of Eq.~\eqref{eq:HH_disp} for the WL hole states.  

The density dependence of the renormalized electron and hole 
 energies is shown in Fig.~\ref{fig:QD-energy} (a) and (c), 
respectively, in the presence of the built-in field. 
The corresponding results without the built-in field are given in
Fig.~\ref{fig:QD-energy} (b) and (d) for comparison. The renormalized
d-shell is split into two degenerate d$_{\pm}$ states and a d$_0$
state, with a separation of a few meV. 

\begin{figure}[tb]
  \begin{center}
  \includegraphics*[width=.45\textwidth]{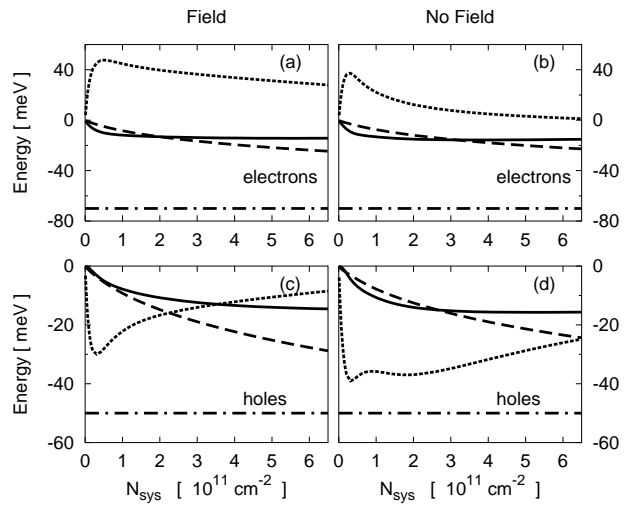}
    \caption{Same as Fig.~\ref{fig:QD-energy} for different contributions to 
 		the renormalized p-shell energy 
                $\widetilde{\varepsilon}_{\text{p}}$
		which are the free energy $\varepsilon_{\text{p}}$ 
                (dashed dotted line), screened Hartree shift  
                $\Delta_{\text{p}}^{\text{SH}}$ 
                (dotted line),
		screened exchange shift $\Delta_{\text{p}}^{\text{SX}}$ 
                (solid line),
		and Coulomb hole shift $\Delta^{\text{CH}}_{\text{p}}$ 
                (dashed line).
             }
    \label{fig:QD-energy-pshell}
  \end{center}
\end{figure}

Quantitatively, the QD levels experience a smaller energy shift for the
built-in field compared to the zero field case. The origin of this
difference lies in the Hartree term, which reflects the electrostatic
interaction of a given carrier with all the others. The 
field-induced change of the $z$-confinement functions tends to 
separate the electrons from the holes and, as a consequence for both,
the repulsive part of the Hartree term is increased and the attractive
part is decreased.   
This effect is illustrated explicitly in Fig.~\ref{fig:QD-energy-pshell}
for the p-shell where the four different contributions to the renormalized
energies are shown.
For electrons the Hartree shift is repulsive both in the
presence and in the absence of the built-in field, but more so in the
former case. For holes, the built-in field makes the Hartree term less
attractive. In both cases the net result is a set of shallower bound 
states. 
For the sake of completeness we mention that
the different sign of the Hartree field for electrons and
holes comes from  the difference in the QD population of electrons and
holes and from the band dependence of the Coulomb matrix elements.
 
The screened exchange and Coulomb hole terms are not significantly
different for with and without built-in field.
As the extended WL states are only renormalized by the screened
exchange and Coulomb hole term, we find an overall negative energy 
shift, lowering the free spectrum by an almost $\V{k}$-independent 
shift (not shown).

%%%%%%%%%%%%%%

%%%%%%%%%%%%%%%%%%%%%%%%%%%%%%%%%%%%%%%%%%%%%%%%%%%%%%%%%%%%%%%%%%%%%%%%%%%%%%%
\subsection{Capture and relaxation times}

To quantify the importance of the different scattering processes, we study their
dependence on the total carrier density in thermal equilibrium.
Using the relaxation time approximation \cite{Nielsen:04} one can introduce a
scattering time $\tau_{\nu}$ for a given process according to 
%%%%%%%%%%%%%%%%
\begin{eqnarray}
 \tau_{\nu} & = & \left[ 
			S_{\nu}^{\text{in}} + S_{\nu}^{\text{out}}
		\right]^{-1} ,
\end{eqnarray}
%%%%%%%%%%%%%%%%
which gives  a characteristic time on which the system will return to 
its thermal equilibrium distribution if exposed to a small perturbation, i.e.,
$\dot{f}_{\nu}=-(f_{\nu}-F_{\nu})/\tau_{\nu}$, with $F_{\nu}$ being the 
thermal equilibrium distribution.

The scattering times are changed by the built-in field through
different competing mechanisms. On the one hand the matrix elements
are modified (see Section~\ref{sec:S_P}), on the other hand the QD 
energies are pushed closer to the WL continuum (Section~\ref{sec:ren}).

\begin{figure}%[h]
  \begin{center}
  \includegraphics*[width=.45\textwidth]{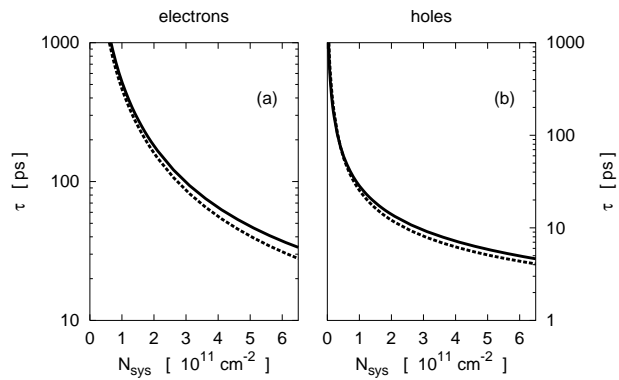}
    \caption{Capture times for the p-shell as a function of the 
		total carrier density in the system $N_{\text{sys}}$
             using the free (unrenormalized energies), with (solid lines) 
		and without (dotted lines) built-in electrostatic field.
		  }
    \label{fig:cap-free}
  \end{center}
\end{figure}

For illustrative purposes, we first study the influence of the built-in field 
on the scattering times by using the free (unrenormalized) energies 
within the scattering integrals.
This reveals the field effect solely on the Coulomb matrix elements 
via the wave-function changes.
As an example we consider a capture process, 
where an electron or hole from the WL is scattered into the QD while another WL 
carrier (electron or hole) is scattered to an energetically higher WL state, 
as well as the reverse process. (Both contribute to the scattering
time according to the relaxation-time approximation.)  
Thus the outer index in Eq.~\eqref{eq:in-out-scatt} 
is a QD state while the three summation indices in
Eq.~\eqref{eq:in-rate} belong to the  WL states for this example
shown in Figure \ref{fig:cap-free}. 

First we discuss the density dependence.
The capture time decreases with increasing carrier density, as 
more scattering partners become available. 
Furthermore, capture times for
electrons are slower than for holes, because the QD electron levels are placed 
energetically deeper below the WL continuum edge. 
For energy-conserving scattering processes, the excess energy of a
WL electron which is captured to the QD 
must be transferred to another carrier from the WL 
in the present example. 
In this way a capture to an 
energetically deep lying QD state is associated with large momentum transfer 
for the WL carriers. As  the matrix elements have a Gaussian dependence on the 
in-plane momentum, scattering to WL states with high momentum is suppressed.
  
%%%%%%%%%%%%%%

The changes in the capture times produced by the field in the case of
unrenormalized energies are minimal. This proves that the competing 
trends described in Section~\ref{sec:S_P} are nearly compensating each 
other, with a slight dominance of the effect of electron-hole
scattering reduction. 

%%%%%%%%%%%%%%
\begin{figure}[tb]
  \begin{center}
  \includegraphics*[width=.45\textwidth]{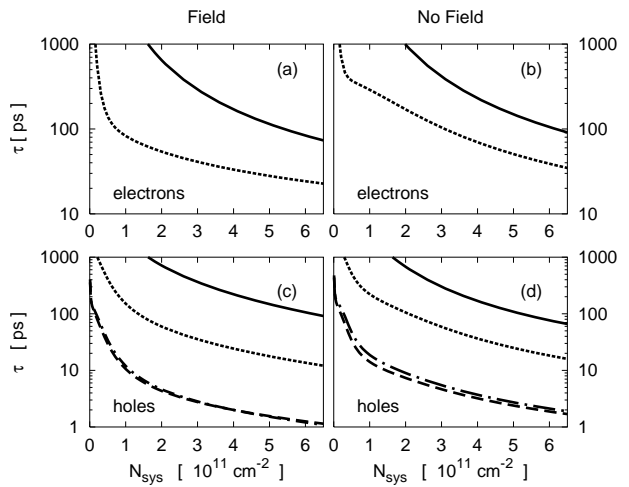}
    \caption{Capture times as a function of the 
		total carrier density in the system $N_{\text{sys}}$
		for s-shell (solid lines), p-shell (dotted lines), 
		d$_0$-shell (dashed-dotted lines), and  
		d$_\pm$-shell (dashed lines).
             }
    \label{fig:capture}
  \end{center}
\end{figure}

For the calculation of the capture times in Fig.~\ref{fig:capture},
based on WL assisted capture processes,\cite{Nielsen:04}
the renormalized energies are included in the scattering integrals 
$S_{\nu}^{\text{in,out}}$. Now the capture times become substantially
shorter in the presence of the built-in field compared to the zero field case.  
Thus, the energy separation of the QD levels from the WL edge plays now the 
dominant role in the scattering times. 
Even though for the built-in field, the capture processes are somewhat 
slowed down by the reduction of the electron-hole interaction, they are still
faster compared to the zero field case, where due to the electron-hole 
interaction the QD levels are energetically deeper in the QD.

\begin{figure}[tb]
  \begin{center}
  \includegraphics*[width=.45\textwidth]{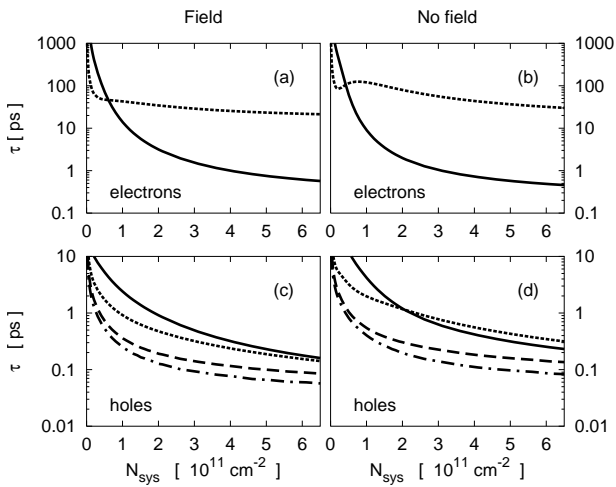}
    \caption{Relaxation times as a function of the 
		total carrier density in the system $N_{\text{sys}}$ with
		labeling as in Fig.~\ref{fig:capture}.
             }
    \label{fig:relaxation}
  \end{center}
\end{figure}

Figure \ref{fig:relaxation} shows the WL assisted QD relaxation times
for processes where a QD electron (hole) scatters to a 
different QD electron (hole) state by means of a WL carrier.
Alternatively, a QD carrier performs a transition to the WL while
another WL carrier scatters into a different QD state.\cite{Nielsen:04}
Thus the outer index in 
Eq.~\eqref{eq:in-out-scatt} belongs to a QD state while two of the summation 
indices in Eq.~\eqref{eq:in-rate} correspond to WL states and one label to a
QD state.  Mixed QD relaxation processes,\cite{Nielsen:04} where
e.g. a QD electron scatters out to the 
WL while another hole from the WL scatters down to the QD hole state
play only a minor role due to the charge separation of the electrons and
holes caused by the built-in field. 

%%%%%%%%%%%%%%
 
Generally, the relaxation times for holes are one or two orders of 
magnitude shorter than for electrons, since the QD energy level spacing is 
larger for the latter. With built-in field the relaxation times become
shorter compared to the zero field case, where the QD energy level
spacing is larger (with the exception of a slightly slower relaxation
of the s-shell electrons). Relaxation times for scattering between
QD states are in general more than one order of magnitude shorter than
capture. For the p-shell electron relaxation, a saturation effect
due to Pauli blocking is observed at higher densities, 
which leads to comparable capture and relaxation times.

%%%%%%%%%%%%%%%%%%%%%%%%%%%%%%%%%%%%%%%%%%%%%%%%%%%%%%%%%%%%%%%%%%%%%%%%%%%%%%%
\section{Conclusion}
The presence of the built-in electrostatic field in wurtzite
heterostructures causes a charge separation of electrons and holes
along the growth direction, which in turn reduces the electron-hole
interaction and increases the electron-electron and hole-hole
interaction. 

Our results show that the usually discussed importance of this effect
on the interaction matrix elements only weakly influences the
scattering rates. It turns out that the change of the
self-consistently renormalized energies due to charge separation leads
to a much stronger modification of the scattering rates.  
Specifically, for the influence on the interaction matrix elements,
the reduction of electron-hole scattering is partly compensated by the
increase of electron-electron and hole-hole scattering. In contrast, 
for the energy renormalization, the charge separation leads to
increased repulsion and decreased attraction in the Hartree terms, 
both effects working in the same direction of shallower confined levels.
This in turn causes an enhancement of the scattering efficiency.  

As expected for the QD-WL system, the rates for carrier capture and
relaxation strongly depend on the density of excited carriers in the
localized and delocalized states. For intermediate densities, the
scattering efficiency increases with carrier density.  For large
densities, Pauli-blocking and screening of the interaction matrix
elements slow down a further increase of the scattering rates. 
For typical InGaN QD parameters, a QD density of 10$^{10}$ cm$^{-2}$
and a carrier density of 10$^{11}$ cm$^{-2}$ at room temperature, direct 
capture of
electrons (holes) to excited QD states results in scattering times on
the order of 100 (10) ps. Relaxation times for scattering between the
QD hole states are more than one order of magnitude shorter than
capture, while at elevated densities the electron p-shell relaxation 
is of the same order of magnitude as capture.

%%%%%%%%%%%%%%%%%%%%%%%
\begin{acknowledgments}
  This work was supported by the Deutsche Forschungsgemeinschaft and
  with a grant for CPU time at the NIC, Forschungszentrum J\"{u}lich.
\end{acknowledgments}

%%%%%%%%%%%%%%%%%%%%%%%%%%%%%%%%%%%%%%%%%%%%%%%%%%%%%%%%%%%%%%%%%%%%%%%%%%%%%%%
\begin{appendix}
\section{OPW states and interaction matrix elements} \label{app:a}

The basic idea of the following scheme is to construct an
approximate single-particle basis for the combined QD-WL system
which provides a feasible way to compute the interaction matrix
elements, Eq.~\eqref{eq:Coulomb}. We start from localized QD
states $ \varphi_\alpha $, as introduced in Sections~\ref{sec:QDmodel}
and \ref{sec:HF} with $\alpha=(m,\V{R})$, and WL states in the
absence of QDs, which are assumed to have plane-wave envelope
functions $\varphi_{\bf k}^0({\bm \varrho})= 1/\sqrt{A} \, e^{i {\bf
k}\cdot {\bm \varrho}} $ in the WL plane with the two-dimensional
carrier momentum $\V{k}$. Quantum numbers for spin, band index, and
confinement in $z$-direction are not explicitly written for
notational simplicity. In the presence of the QDs the orthogonality
condition of the basis is imposed by projecting the plane waves on the 
subspace orthogonal to the QD states, as outlined in
Ref.~[\onlinecite{Nielsen:04}]. 
The WL states are therefore given by the OPW functions
$ | \varphi_{\bf k} \rangle = \frac{1}{N_{\bf k}} \: ( |
\varphi_{\bf k}^0 \rangle - \sum_\alpha | \varphi_\alpha \rangle
\langle \varphi_\alpha \: | \varphi_{\bf k}^0 \rangle ) $. 
Assuming QDs with nonoverlapping wave functions, the sum over
$\alpha=(m,\V{R})$ counts various QD states $m$ at different QD
positions $\V{R}$.  For randomly distributed identical QDs, the
normalization is given by $ N_{\bf k}^2 =1-N \sum_m|\langle \varphi_m |
\varphi_{\bf k}^{0} \rangle|^{2}$.

This scheme allows to evaluate the in-plane integrals $ \left\langle
\nu | e^{i \V{q} \cdot {\bm\varrho} } | \nu' \right\rangle = \int
\mathrm{d}^2 \varrho \; \varphi_{\nu}^{\ast}( {\bm \varrho} ) e^{i
\V{q} \cdot {\bm\varrho} } \varphi_{\nu'}( {\bm\varrho} ) $ which
appear in the Coulomb matrix elements, Eq.~\eqref{eq:Coulomb2}, for
various combinations of QD and WL states.  When $\nu$ and $\nu'$ are
two QD states, one obtains
%%%%%%%%%%%%%%%%
\begin{align}
\left\langle \; \alpha  \; | e^{i \V{q} \cdot {\bm\varrho} } | 
             \; \alpha' \; \right\rangle 
 & =                                                              \label{eq:dd}
 \left\langle \; m  \; | e^{i \V{q} \cdot {\bm\varrho} } | 
             \; m' \; \right\rangle
 \; e^{i \V{q} \cdot \V{R}} \; \delta_{\V{R},\V{R}'}           \, ,
\end{align}
%%%%%%%%%%%%%%%%
with the QD positions $\V{R}$ and $\V{R}'$. For combinations of QD and
WL states, one finds
%%%%%%%%%%%%%%%%
\begin{align}
\left\langle \; \alpha  \; | e^{i \V{q} \cdot {\bm\varrho} } | 
             \; \varphi_{{\bf k}'} \; \right\rangle 
 & =                                                              \label{eq:dk}
\left\langle \; m  \; | e^{i \V{q} \cdot {\bm\varrho} } | 
             \; \varphi_{{\bf k}'} \; \right\rangle 
 \; e^{i (\V{k}' + \V{q}) \cdot \V{R}}      \, ,
\end{align}
%%%%%%%%%%%%%%%%
and for two WL states 
%%%%%%%%%%%%%%%%
\begin{align}
\left\langle \; \varphi_{\bf k}     \; | e^{i \V{q} \cdot {\bm\varrho} }  | \; 
                \varphi_{{\bf k}'}  \; 
 \right\rangle \label{eq:kk}
  & =  
 \delta_{\V{k}, \, \V{q}  +\V{k}'} \;  
 D_{\text{OPW}}(\V{k},\V{k}',\V{q})        \, ,
\end{align}
%%%%%%%%%%%%%%%%
follows with
%%%%%%%%%%%%%%%%
\begin{align}
  D_{\text{OPW}}&(\V{k},\V{k}',\V{q}) =   
 \frac{1}{N_{\V{k}}N_{\V{k}'}}             
\label{eq:delta_opw} \\ 
\times & [ 1 - N \sum \limits_m |
     \langle \varphi_{\bf k}^0 | \varphi_m \rangle |^2 
     - N \sum \limits_m 
     | \langle \varphi_m |\varphi_{{\bf k}'}^0 \rangle |^2 
\nonumber \\ 
   &  +  N
    \sum \limits_{m,m'}
     \langle \varphi_{\bf k}^0 | \varphi_m \rangle 
     \langle \varphi_m | e^{i {\bf q}\cdot{\bm \varrho}} | \varphi_{m'} \rangle
     \langle\varphi_{m'} | \varphi_{{\bf k}'}^0 \rangle ].
\nonumber
\end{align}
%%%%%%%%%%%%%%%%

The orthogonality requirement of the wave functions is directly
reflected in the interaction vertices $\left\langle \nu_1| e^{+i \V{q}
\cdot {\bm\varrho} } | \nu_2\right\rangle = \delta_{\nu_1, \nu_2} $
for \mbox{$\V{q}=0$}.  Meaningful results for the interaction matrix
elements can be expected only when the approximate model shows
the same behavior. Since we start from orthogonal QD states at a given
QD position and assume nonoverlapping wave functions for different
QDs, Eq.~\eqref{eq:dd} reduces to a Kronecker delta for
\mbox{$\V{q}=0$}.  QD and OPW-WL states are orthogonal by
construction, i.e., the requirement is also fulfilled for
Eq.~\eqref{eq:dk}.  As described in Ref.~[\onlinecite{Nielsen:04}], it is the
assumption of randomly distributed QDs which, in the large-area limit
restores - on average - translational invariance and provides mutually
orthogonal OPW states such that the above requirement is also obeyed
by Eq.~\eqref{eq:kk}. Note that $ D_{\text{OPW}}(\V{k},\V{k},0)=1$.

%%%%%%%%%%%%%%%%%%%%%%%%%%%%%%%%%%%%%%%%%%%%%%%%%%%%%%%%%%%%%%%%%%%%%%%%%%%%%%%
\section{Hartree energy renormalization} \label{app:b}

Starting from Eq.~\eqref{eq:sigma_hf}, the Hartree energy shift of the
QD states has contributions form the QD and from the WL carriers,
%%%%%%%%%%%%%%%%
\begin{align} 
 \Delta^{\text{H}}_{b,\alpha} & = 
  \sum_{b',\alpha'} V^{b,b'}_{\alpha \alpha' \alpha' \alpha} 
   \; f^{b'}_{\alpha'}
 +
  \sum_{b',\V{k}'} V^{b,b'}_{\alpha \V{k}' \V{k}' \alpha} 
   \; f^{b'}_{\V{k}'} 
 \notag \\
 & = \Delta^{\text{H}, \text{QD}}_{b,\alpha} +
    \Delta^{\text{H}, \text{WL}}_{b, \alpha} \; .
\label{eq:H} 
\end{align}
%%%%%%%%%%%%%%%%%
The QD contribution can be specified further using the notation
introduced above and Eq.~\eqref{eq:Coulomb2},
%%%%%%%%%%%%%%%%
\begin{align}
 \Delta^{\text{H}, \text{QD}}_{b, \alpha}& =
 \sum_{\V{R}'}\sum_{b',m'} \frac{1}{A} \sum_{\V{q}} 
  V^{b,b'}(\V{q})  \;  f_{m'}^{b'}
  \notag \\
  & \times 
  \langle m  |  e^{-i \V{q} \cdot {\bm \varrho} } | m  \rangle 
  \langle m' |  e^{i \V{q} \cdot {\bm \varrho} } | m' \rangle
  \notag \\
  & \times 
   e^{-i \V{q}\cdot( \V{R} - \V{R}')} .
\end{align}
%%%%%%%%%%%%%%%%%
The result depends on the QD positions through the
phase factor arising from the interaction vertices as given by
Eqs.~\eqref{eq:dd}-\eqref{eq:kk}. In the large area limit one
obtains, by the law of large numbers, that the distribution of these
quantities is sharply peaked around their configurational averaged 
value. Therefore we may replace the QD contribution in Eq.~\eqref{eq:H} by
%%%%%%%%%%%%%%%%
\begin{align}
 \Delta^{\text{H},\text{QD}}_{b,m}& = 
\frac{1}{N} \sum_{\V{R},\V{R}'}\sum_{b',m'} \frac{1}{A} \sum_{\V{q}} 
  V^{b,b'}(\V{q})  \;  f_{m'}^{b'}
  \notag \\
  & \times 
  \langle m  |  e^{-i \V{q} \cdot {\bm \varrho} } | m  \rangle 
  \langle m' |  e^{ i \V{q} \cdot {\bm \varrho} } | m' \rangle
  \notag \\
  & \times 
   e^{-i \V{q}\cdot( \V{R} - \V{R}')} . 
\label{eq:HQD}
\end{align}
%%%%%%%%%%%%%%%%%
Note that the resulting Hartree shift does not depend on the QD
position any more. The summation over the random
positions $\V{R}, \V{R}'$ is evaluated as in the disordered system
theory (see e.g. Ref.~\cite{Doniach:98}):
%%%%%%%%%%%%%%%%
\begin{align}
\label{eq:average}
\sum_{\V{R}, \V{R}'} f(\V{R})g( \V{R}') & = \sum_{\V{R} \neq \V{R}'}
f(\V{R})g( \V{R}') + \sum_{\V{R}}f(\V{R})g( \V{R}) \notag \\
 & = N^2 \lav f \rav \cdot  \lav g \rav + N \lav fg\rav ,
\end{align}
%%%%%%%%%%%%%%%%%
where $\lav F \rav = 1/A \int d^2 R \, F(\V{R})$ denotes the
configuration average. The first term is the uncorrelated
average of the two random variables, while the second takes into
account that for the same point they are correlated.

In our case $f(\V{R})= e^{-i \V{q} \cdot \V{R}},g(\V{R}')= e^{i \V{q}
  \cdot \V{R}'}$, $ \lav f \rav = \lav g \rav=\delta_{\V{q},0}$ and
$ \lav f g \rav = 1$, so that one may write:
%%%%%%%%%%%%%%%%
\begin{align}
  \Delta^{\text{H},\text{QD}}_{b,m}& = 
  n_{\text{QD}} \sum_{b',m'}  
  V^{b,b'}(\V{q}=0)   
  f_{m'}^{b'} \notag \\
  &+ \sum_{b',m'} V_{m m' m' m}^{b,b'} 
  \; f_{m'}^{b'}.
\label{eq:Hartree1}  
\end{align}
%%%%%%%%%%%%%%%%%  
The first term, arising from the Coulomb interaction between different
QDs is proportional to the total QD charge density, while the second 
one describes the Hartree interaction inside a given dot.     

The WL contribution to the QD Hartree energy shift can be evaluated 
similarly,
%%%%%%%%%%%%%%%%
\begin{eqnarray}
 \Delta^{\text{H}, \text{WL}}_{b,\alpha} & = & 
 \sum_{b',\V{k}'} \frac{1}{A} \sum_{\V{q}} 
 V^{b,b'} (\V{q}) \; f_{\V{k}'}^{b'}
  \notag \\
  & \times &
  \langle m  |  e^{-i \V{q} \cdot {\bm \varrho} } | m  \rangle 
   e^{-i \V{q} \cdot \V{R}}    \notag \\
  & \times &    
  \delta_{\V{k}',\V{k}'+\V{q}} D_{\text{opw}}(\V{k},\V{k}',\V{q}) 
 \notag \\
  &=&  \frac{1}{A} \sum_{b',\V{k}'}
  V^{b,b'} (\V{q}=0) f_{\V{k}'}^{b'} .
\label{eq:Hartree2} 
\end{eqnarray}
%%%%%%%%%%%%%%%%%
The result has the same structure as the first term of Eq.~\eqref{eq:Hartree1}
and adds the WL charge density to the QD contribution. By charge neutrality
these terms cancel each other and, as expected, the Coulomb
singularity at $\V{q}=0$ is removed. One is left with the second term of
Eq.~\eqref{eq:Hartree1}, which gives Eq.~\eqref{eq:H_dot}.

Following the same steps as above we find for the  WL Hartree energy shift
%%%%%%%%%%%%%%%%
\begin{align} 
 \Delta^{\text{H}}_{b, \V{k}} & = 
  \sum_{b',\alpha'} V^{b,b'}_{\V{k} \alpha' \alpha' \V{k}} 
   \; f^{b'}_{\alpha'}
 +
  \sum_{b',\V{k}'} V_{\V{k} \V{k}' \V{k}' \V{k}}^{b,b'} 
f_{\V{k}'}^{b'}  \notag
 \\
 & =  V^{b,b'} (\V{q}=0)
 \Big\{
   n_{\text{QD}}\sum_{b',m'}  f_{m'}^{b'}
 + 
   \frac{1}{A}\sum_{b', \V{k}'} f_{\V{k}'}^{b'} 
 \Big\} \notag
 \\ 
 & =  0 , 
\end{align}
%%%%%%%%%%%%%%%%%
which vanishes due to global charge neutrality.

%%%%%%%%%%%%%%%%%%%%%%%%%%%%%%%%%%%%%%%%%%%%%%%%%%%%%%%%%%%%%%%%%%%%%%%%
\section{WL screening contributions to the QD Hartree interaction}
\label{app:c} 

In our description of the Hartree terms in the previous Appendix~\ref{app:b}, 
the summation over randomly
distributed QDs restores, in the large area limit, the in-plane
translational invariance of the OPW-WL states. On this level, only the
averaged QD properties enter - a picture which is consistent with the
expectation that in a system with a macroscopic number of QDs, like a
QD-laser, only the averaged properties of the QD ensemble should be
important. On a local scale at a QD position, however, the WL states
do not obey translational invariance since perturbations of the WL
states due to the QD appear. 
In truly homogeneous systems one has a $\V{q}=0$ Coulomb singularity 
which is canceled out by the global charge neutrality and no other 
Hartree contribution is present. 
We have shown in Appendix~\ref{app:b} for the system of randomly distributed 
QDs on the WL,
%
%Here one expects 
that a similar 
singularity is produced by the configuration averaging and is removed 
by global neutrality arguments.
%, but local Hartree fields are still
%felt by the carriers in the QD. 
Then only the localized QD carriers are subjected to Hartree fields induced 
by carriers in the same QD.

In this appendix, we reexamine the result using the Green's function (GF)
formalism.\cite{Doniach:98} The more refined treatment shows that WL carriers 
can provide corrections to the QD Hartree shift which can be cast into the form
of screening contributions.

\begin{figure}[tb]
    \begin{center}
    \includegraphics*[width=.45\textwidth]{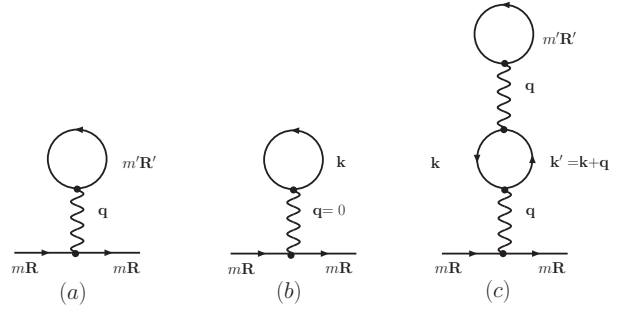}
    \caption{Hartree diagrams discussed in the text. \label{dgrm}}
    \end{center}
%    \label{dgrm}
\end{figure}

%We illustrate the phenomenon, using the Green's function (GF)
%formalism.\cite{Doniach:98} 
Some low order diagrams in the GF expansion,
describing terms of the Hartree contribution to the QD energies,
are shown in Fig.~\ref{dgrm}. Since the QDs are identical, the QD 
propagators are position independent, but the interaction vertices
contain phase factors related to the position and to the adjoining
momenta, as given by  Eqs. (\ref{eq:dd})-(\ref{eq:kk}). The procedure
described in the previous Appendix amounts to the averaging of the
Hartree self-energies. 

To begin with, the self-energy in diagram (a) has the form
\begin{align}
-i\hbar &\frac{1}{N} \sum_{\V{R},\V{R}'}\sum_{b',m'} \frac{1}{A} \sum_{\V{q}} 
  V^{b,b'}(\V{q})  \;  G_{m'}^{b'}(t,t)
  \notag \\
  & \times 
  \langle m  |  e^{-i \V{q} \cdot {\bm \varrho} } | m  \rangle 
  \langle m' |  e^{ i \V{q} \cdot {\bm \varrho} } | m' \rangle
  \notag \\
  & \times e^{-i \V{q}\cdot( \V{R} - \V{R}')} \; .
\end{align}
In a self-consistent calculation, with the equal-time
GF related to the population factors in the usual way, this leads
to the QD Hartree term of Eq. (\ref{eq:HQD}). In the large area
limit, using Eq. (\ref{eq:average}), one obtains the two terms of
Eq. (\ref{eq:Hartree1}). As noted before, the first term corresponds
to averaging the 'tadpole head' of diagram (a) as if it would be
independent of the rest of the diagram, while the second term contains
the contributions of the correlations.
It is easy to see that diagram (b) of Fig.~\ref{dgrm} leads to the WL 
Hartree contribution spelled out in Eq. (\ref{eq:Hartree2}). This term,
together with the first term of diagram (a) contain the Coulomb 
$\V{q}=0$ singularity specific to homogeneous systems. The second term
of diagram (a) is an example of a local field which is not
averaged out, the source of this field being 'in phase' with the
charges that probe it.
   
%%%%%%%%%%%%%%%%%%
%%%%%%%%%%%%%%%%%%

A similar analysis  is valid for diagram (c). Its phase factors are
the same as in diagram (a) and again one has two contributions. The
first one, coming from the uncorrelated averaging, leads to a
self-energy containing only the $\V{q}=0$ contribution, 
\begin{align}
(i \hbar)^2 \frac{N}{A^2}&
  \; V^{b,b_1}(\V{q}=0)G_{\V{k}}^{b_1}(t,t')G_{\V{k}}^{b_1}(t',t)
  \notag \\ 
 \times & \;V^{b_1,b'}(\V{q}=0)G_{m'}^{b'}(t',t') ,
\end{align}
where summation and integration over the inner variables is assumed. 
Since this entails the restriction $\V{k}=\V{k}'$, the diagram
contributes to the renormalization of the WL propagator of index
$\V{k}$ in diagram (b). Therefore this is already included in a
self-consistent calculation. 
More interesting is the second term, arising from the correlated
averaging, 
\begin{align}
(i \hbar)^2 \frac{1}{A^2}&
  \;
  V^{b,b_1}_{m,\V{k},\V{k}+\V{q},m} G_{\V{k}+\V{q}}^{b_1}(t,t')
  G_{\V{k}}^{b_1}(t',t)    
  \notag \\ 
 \times & \;V^{b_1,b'}_{\V{k}+\V{q},m',m',\V{k}} G_{m'}^{b'}(t',t') .
\end{align}
In this case the summation 
over the momentum transfer $\V{q}$ remains unrestricted.
The structure is similar to the second term of diagram (a), i.e., it
corresponds to  the intra-QD Hartree field, but with the additional
$\V{k}$ and $\V{k}'=\V{k}+\V{q} $  WL propagators forming a Lindhard
loop. The loop describes the screening of the
intra-QD Hartree field by the WL carriers. 

The usual procedure in the GF theory is to leave the Hartree
interaction unscreened and to add the Lindhard loop of diagram (c)
to the 'tadpole head' GF of the diagram (b).
This avoids the double counting of such diagrams. As a result, a nondiagonal
$(\V{k},\V{k}')$ GF appears in the self-consistent Hartree loop of 
diagram (b). Alternatively, one can avoid double counting by keeping 
only momentum-diagonal WL propagators and leave the Lindhard loop 
for the screening of the Coulomb line. We have chosen this second
approach, which also considerably simplifies the formalism.

A fully systematic analysis of all the possible diagrams and the
action of the configuration averaging over them is way beyond the 
scope of this paper. The approximation proposed here includes the 
following physically important features. The random
phases associated with the QD positions give rise to a $\V{q}=0$
singularity, which is canceled out by the global charge neutrality. On
the other hand, the intra-QD fields are not influenced by the phase
factors and therefore are not averaged out. The same is true for the
local WL charges that respond to these fields and induce their
screening.

\end{appendix}

%\bibliography{refs}
%%%%%%%%%%%%%%%%%%%%%%%%%%%%%%%%%%%%%%%%%%%%%%%%%%%%%%%%%%%%%%%%%%%%%%%%%%%%%%%

\end{document}